\documentclass{aa}  
\usepackage{graphicx}
\usepackage{txfonts}
\usepackage{color}
\usepackage[utf8]{inputenc}
\usepackage{chngcntr}
\usepackage{hyperref}
\hypersetup{%draft,
  citecolor=cyan,      % color of links to bibliography
  filecolor=magenta,      % color of file links
  urlcolor=magenta            % color of external links
}

\makeatletter
\renewcommand{\thefigure}{A\@arabic\c@figure}
\@addtoreset{figure}{section}
\makeatother

\begin{document} 

   \title{Identifying Long Radio Transients with Accompanying X-Ray   \\
  Emission as Disk-Jet Precessing Black Holes: The Case of ASKAP J1832-0911}

   %\subtitle{I. Overviewing the $\kappa$-mechanism}

   \author{Antonios Nathanail\thanks{anathanail@academyofathens.gr}}
%           \inst{1,2},%\fnmsep  ,
%           Yosuke Mizuno\inst{3,4,2}, 
%            Ioannis Contopoulos\inst{1}, 
%            Christian M. Fromm\inst{5,2,6},  Alejandro Cruz-Osorio\inst{7}, \\
%   Kotaro Moriyama\inst{8,2}  \and Luciano Rezzolla\inst{2,9,10}  }
% % List of institutions

   \institute{Research Center for Astronomy and Applied Mathematics,
 Academy of Athens, Athens 11527, Greece} %\\

   \date{Received June 18, 2025; Accepted --- 2025}

% \abstract{}{}{}{}{} 
% 5 {} token are mandatory
 
  \abstract
  % context heading (optional)
  % {} leave it empty if necessary  
   {}
  % aims heading (mandatory)
   {In this work we investigate whether the 2 min bursts every 44 min from 
   ASKAP J1832-0911 can be explained by Lense-Thirring 
   precession of an intermediate-mass black hole (IMBH) 
   accretion disk launching a Blandford-Znajek jet, as 
   an alternative to magnetar or white-dwarf models.}
  % methods heading (mandatory)
   {We derive the Lense-Thirring period 
\( P_{\rm LT}=\frac{\pi G M}{a c^3}r^3 \)
and solve \(P_{\rm LT}=44\) min for black-hole mass
\(M\) and dimensionless radius \(r=R/R_g\).  
We estimate the equipartition field \(B\) at 
\(r\), compute the Blandford-Znajek power
\( P_{\rm BZ}, \), and the power expected from 
a gap at the black hole magnetosphere,
and compare the resulting jet luminosity to the 
observed radio and X-ray fluxes at \(D\approx4.5\) kpc.  
We also evaluate expected high-frequency variability 
and the angular size for Very Long Baseline Interferometry (VLBI)
observations.}
  % results heading (mandatory)
   {For \(a\sim0.3\!-\!0.9\), an IMBH with 
   \(M\sim10^3\!-\!10^5\,M_\odot\) yields 
   \(r\sim10\!-\!40\,R_g\) and \(P_{\rm LT}=44\) min.  
   Equipartition gives \(B\sim10^5\) G at \(r\), 
   leading to \(P_{\rm BZ}\sim10^{35\!-\!39}\) erg ${\rm s^{-1}}$. 
   With radiative efficiency 
   \(\epsilon_j\sim10^{-2}\!-\!10^{-1}\), the predicted 
   \(L_{\rm jet}\sim10^{34\!-\!36}\) erg ${\rm s^{-1}}$ matches the 
   observed \(F_X\sim10^{-12}\) erg ${\rm cm^{-2}}$ ${\rm s^{-1}}$ and radio flux,
   variability on \(\lesssim100\) s could be a smoking gun of this model.}
%and the jet's \(\lesssim10\,\mu\)as 
% size requires space-VLBI at \(\ge86\) GHz for resolution.}
  % conclusions heading (optional), leave it empty if necessary 
   {The IMBH precessing-jet model simultaneously explains 
   the periodicity, energetics, and duty cycle of ASKAP 
   J1832-0911.  Only high-time-resolution X-ray timing 
   (to exclude \(\sim\)s pulsations) and multi-frequency 
   radio polarimetry (to confirm a flat, low-polarization 
   spectrum) can definitively distinguish it from magnetar 
   or white-dwarf scenarios.}

   \keywords{black hole physics --
   resistivity -- accretion -- accretion discs -- 
magnetic reconnection  -- magneto-hydrodynamics
}

\titlerunning{Long Radio Transients as Precessing IMBH}
\authorrunning{Nathanail Antonios}
    %\lhead{Nathanail et al.}
    %\rhead{Resistivity and variability}
   \maketitle
%
%-------------------------------------------------------------------

\section{Introduction}
Long radio transients
-sources whose emission appears 
abruptly and fades over timescales ranging from minutes 
to days-have long challenged our understanding of 
compact-object astrophysics\citep{Hurley2022, Hurley2023}.
A handful of Galactic 
Center Radio Transients (GCRTs), such as GCRT J1745-3009
with its ~10 min bursts every ~77 min, exemplify periodic
behavior on timescales of tens of minutes \citep{Hyman2005}. 
More recently,
ASKAP J1832-0911 was discovered as a bright radio-and X-ray 
transient located in the Scutum spiral arm at an 
estimated distance of ~4.5 kpc. ASKAP observations reveal
~2 min-long radio flares (at GHz frequencies, with peak 
flux densities of order 10 mJy) recurring every 44 min, 
and simultaneous X-ray observations detect 2 min X-ray 
pulses (with fluxes $~10^{-12}\,\, {\rm erg\,\, cm^{-2}\,\, s^{-1}}$ 
 in 0.5-10 keV) 
at the same 44 min cadence \citep{Wang2025}. 
Such a precise, multiwavelength 
periodicity over tens of minutes is exceedingly
rare among known transients.

Two leading compact-object interpretations are followed: 
a highly magnetized neutron star (magnetar) whose beamed emission 
sweeps past Earth for ~2 min each 44 min rotation \citep{Cooper2024}, and 
an ultracompact white-dwarf (WD) binary in which magnetic 
gating produces synchronized radio and X-ray bursts 
every orbital period. 
A slowly 
rotating magnetar with occasional "heartbeat" flares 
can reproduce similar with the observed X-ray energetics  \citet{Pons2019}, 
while   analogous ~hour-period WD-WD binaries as 
possible progenitors of long radio flares \citep{Qu2025} with 
X-rays \citep{Schwope2023}, or a WD with a pulsar
\citep{Katz2022}. Although 
these models capture some aspects of the 44 min duty 
cycle, each faces challenges-e.g., maintaining a 44 
min spin despite rapid magnetar spin-down, or explaining
coherent X-ray emission from a WD shock engine 
over minute-long intervals.

Here, we propose instead that ASKAP J1832-0911 is powered 
by an intermediate-mass black hole (IMBH, $M ~ 10^3 - 10^5 \, 
M_{\odot}$) with a tilted accretion disk whose inner regions 
rigidly precess via the Lense-Thirring effect. In this picture, 
a relativistic jet, launched by Blandford-Znajek extraction of 
spin energy, sweeps across our line of sight for ~2 min every 
44 min, producing the observed radio and X-ray flares. IMBHs 
have been invoked to explain ultraluminous X-ray sources 
(ULXs; e.g., \citet{Farrell2009}), dynamical signatures in
globular clusters (e.g., \citet{Gebhardt2005}), and low-luminosity 
AGN in dwarf galaxies  \citep{Mezcua2017}. Formation channels include 
direct collapse of Population III stars, runaway mergers in dense
star clusters \citep{Portegies_Zwart_2002}, and the cores
of dwarf galaxies \citep{Reines2015}. If an IMBH of 
$M~10^5\, M_{\odot}$ resides in a Galactic plane cluster or 
an unrecognized dwarf satellite, its inner disk could 
naturally precess on a ~44 min timescale at radii r~$5-30 \, R_g$, 
while accumulating sufficient magnetic flux to 
power the observed jet luminosity.

This Letter is organized as follows. In Section 2, we 
outline a simple geometric model for Lense-Thirring 
precession of a tilted ring or disk, derive the relationship 
between black-hole mass M, disk radius $r(= R/R_g)$, 
and precession period, and show how the 44 min cycle 
constrains M and r. In Section 3, we apply the model to 
derive viable IMBH mass ranges and disk radii, estimate the 
equipartition magnetic field at r and the corresponding 
Blandford-Znajek jet power, and compare these predictions 
to the observed radio and X-ray luminosities. We also discuss 
 known AGN jet precession as a sanity check on our IMBH scaling
 (e.g. M87's ~11 yr cycle). 
In Section 4, we contrast the IMBH scenario with magnetar
and white-dwarf alternatives by summarizing key observational 
discriminants (e.g., X-ray spectral components, spin pulsations, 
radio polarization) and show why a stellar-mass XRB with 
subhour precession is effectively ruled out by analogues such 
as SS 433, Her X-1, and LMC X-4. We conclude with suggestions 
for targeted follow-up observations to definitively distinguish 
among these models.

%-----------------------------------------------------------------------
% \section{Resistive GRMHD simulations}
% \label{sec:main}
%-----------------------------------------------------------------------
\section{Lense-Thirring Precession and Visibility Geometry}
\label{sec:LTP}

To assess whether a tilted, rigidly precessing disk (and its jet) 
can naturally produce 2 min-long flares every 44 min, we first 
derive the Lense-Thirring (LT) precession period for a test 
ring at radius \(R\) around a spinning black hole.  We then 
construct a simple geometric model \citep{Maccarone2002,
Liska2018,Liska2019a} and relate the jet's half-opening 
angle and disk tilt to the observed duty cycle.

\subsection{Lense-Thirring Precession Period}

A Kerr black hole of mass \(M\) and dimensionless spin \(a\) has angular momentum
$ J \;=\; a\,\frac{G M^2}{c}\, $.
A test particle (or narrow ring) at Boyer-Lindquist 
radius \(R\) experiences a nodal precession frequency 
\(\Omega_{\rm LT}(R)\) given by \citep{Wilkins1972,
 Stella1998,Fragile2007a}:
\begin{equation}
\Omega_{\rm LT}(R)
=\frac{2\,G\,J}{c^2\,R^3}
=\frac{2\,a\,G^2\,M^2}{c^3\,R^3}\,.
\end{equation}
Hence the LT precession period is
\begin{align}
P_{\rm LT}(R)
&=\frac{2\pi}{\Omega_{\rm LT}(R)}
=\frac{2\pi}{\displaystyle \frac{2\,a\,G^2\,M^2}{c^3\,R^3}}
=\frac{\pi\,c^3\,R^3}{a\,G^2\,M^2}\,.
\label{eq:PLT_R}
\end{align}
Measuring distance in gravitational-radius units,
\begin{equation}
R_g = \frac{G M}{c^2}, 
\quad
r \equiv \frac{R}{R_g},
\end{equation}
and substituting \(R = r\,R_g\) into Eq.~\eqref{eq:PLT_R} gives
\begin{equation}
P_{\rm LT}(r)
= \frac{\pi\,G\,M}{a\,c^3}\;r^3
\;=\;
\bigl(1.55\times10^{-5}\bigr)\,\frac{m}{a}\;r^3\quad[\mathrm{s}],
\label{eq:PLT_r}
\end{equation}
where \(M = m\,M_\odot\) and \(G\,M_\odot/c^3\approx4.9\times10^{-6}
\,\mathrm{s}\).  Requiring \(P_{\rm LT}=44\,\mathrm{min}=2640\,
\mathrm{s}\) yields the constraint
\begin{equation}
m\,r^3 = 1.703\times10^8\,a.
\label{eq:mass_radius}
\end{equation}

\subsection{Visibility Geometry and Duty Cycle}

If the inner disk at radius \(r\) is tilted by angle \(\psi\) 
relative to the black-hole spin axis, it precesses at
\(\Omega_{\rm LT}\).  
Assuming the jet is launched along the disk normal 
\citep{Maccarone2002,Fragile2007a,Liska2018}, and that our line
of sight is inclined by \(i\) to the spin axis, the 
instantaneous angle \(\theta(t)\)\footnote{Using the relation of this 
angle from  the dot product of the line of sight and 
the disk normal unit vectors and allowing  the disk normal to 
precess by an azimuthal angle $\phi = \Omega{\rm LT}\,t$, we 
arrive at Eq.~\ref{eq:theta_t}.}
between jet axis and
observer satisfies \citep{Caproni2006,Maccarone2002}
\begin{equation}
\cos\theta(t)
= \cos\psi\,\cos i
+ \sin\psi\,\sin i\;\cos\bigl(\Omega_{\rm LT}\,t\bigr).
\label{eq:theta_t}
\end{equation}
For a jet with half-opening angle \(\alpha\), it is visible
whenever \(\theta(t)\le\alpha\).  Defining \(\Delta t\) as
the total "on" time per cycle \(P_{\rm LT}\), one finds 
at the entry/exit points \(\theta=\alpha\):
\begin{equation}
\cos\alpha
= \cos\psi\,\cos i
+ \sin\psi\,\sin i\;\cos\!\Bigl(\pi\,\frac{\Delta t}{P_{\rm LT}}\Bigr).
\label{eq:alpha_duty}
\end{equation}
Setting \(\Delta t=2\) min and \(P_{\rm LT}=44\) min (so 
\(\pi\Delta t/P_{\rm LT}=\pi/22\)) shows that even modest
tilts \(\psi\sim10^\circ\!-\!30^\circ\) and inclinations 
\(i\sim30^\circ\!-\!70^\circ\) yield small 
\(\alpha\sim5^\circ\!-\!15^\circ\), reproducing the 
2 min/44 min duty cycle \citep[cf.][]{Maccarone2002,Liska2019a}.

In Fig.~\ref{jet} we present a schematic of the tilted 
accretion disk, its precessing jet cone, and the viewing
geometry. An animation illustrating a disk tilt of 
$\psi =20.6^{\circ}$, jet half-opening angle 
$\alpha=10^{\circ}$,  and observer inclination
$i=30^{\circ}$, showing how the cone sweeps across our line of sight every 44 min, is available \href{https://zenodo.org/records/15676045?token=eyJhbGciOiJIUzUxMiJ9.eyJpZCI6ImFlOWQ4NDAwLTdkODAtNGMwOS1iYjkwLTIxYTEwMjA3OWZhNSIsImRhdGEiOnt9LCJyYW5kb20iOiI2YmM2MmYzNGE4NTMyNmI0YjA0NzcyYWUxNjMyNGEyMCJ9.GlfK3o-Djmfa_kINqbq4gMPbUONel_EtJkOJ5uWF7lcSRDfBQMBEJHgez7g26oLtXjsoeLk73-NW0QM7q-EmKw}{here}.

\setcounter{figure}{0}
\renewcommand{\thefigure}{\arabic{figure}}

\begin{figure}%[h]
\centering
\includegraphics[width=0.485\textwidth]{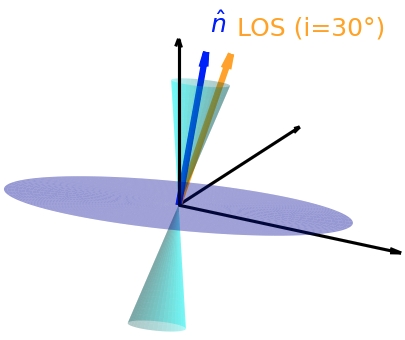}
\caption{Geometry of the precessing disk-jet system. 
The blue disk is tilted by $\psi$ relative to the black 
hole spin axis (vertical), and the cyan cones show 
the jet with half-opening angle $\alpha$. As the system 
precesses, the disk normal (blue arrow) periodically 
aligns with the observer's line of sight (orange 
arrow) and the jet beam crosses the LOS, 
producing the observed 2 min flares every 44 min. An accompanying animation is available  \href{https://zenodo.org/records/15676045?token=eyJhbGciOiJIUzUxMiJ9.eyJpZCI6ImFlOWQ4NDAwLTdkODAtNGMwOS1iYjkwLTIxYTEwMjA3OWZhNSIsImRhdGEiOnt9LCJyYW5kb20iOiI2YmM2MmYzNGE4NTMyNmI0YjA0NzcyYWUxNjMyNGEyMCJ9.GlfK3o-Djmfa_kINqbq4gMPbUONel_EtJkOJ5uWF7lcSRDfBQMBEJHgez7g26oLtXjsoeLk73-NW0QM7q-EmKw}{here}.}
\label{jet}
\end{figure}

%!!!!!!!!!!!!!!!!!!!!!!!!!!!!!!!!!!!!!!!!!!!!!!!!!!!!!!!!!!!!!!!
\section{IMBH Mass Constraints, Jet Power, and Variability}
\label{sec:IMBH}
%--------------------------------------------------------------

Using the Lense-Thirring precession relation 
(Eq.~\ref{eq:mass_radius}),
% \begin{equation}
% m\,r^3 \;=\; 1.706\times10^8\,a,
% \end{equation}
we show that an IMBH (\(M\sim10^3\!-\!10^5\,M_\odot\))
naturally reproduces \(P_{\rm LT}=44\) min at radii 
\(r\sim5\!-\!30\).  We then estimate the magnetic field 
strength at \(r\), compute the Blandford-Znajek jet power, 
compare it to the observed radio/X-ray luminosities 
(for \(D\simeq4.5\) kpc), and discuss small-scale variability 
and VLBI-scale imaging.

\subsection{Mass-Radius Solutions for \(P_{\rm LT}=44\) min}

We need a precession with a period of ~ 44 min, so we plug
\(P_{\rm LT}=2640\) s into Eq.~\eqref{eq:PLT_r} 
and assume \(a\sim0.3\!-\!0.9\), which yields
$r \;=\;\Bigl(\tfrac{1.703\times10^8\,a}{m}\Bigr)^{1/3}.$
and to check where the rigid precession ring is we find:
\begin{itemize}
  \item \(m=10^3\), \(a=0.5\): \(r=(8.53\times10^4)^{1/3}\approx44\).
  \item \(m=10^4\), \(a=0.5\): \(r=(8.53\times10^3)^{1/3}\approx20.4\).
  \item \(m=10^5\), \(a=0.5\): \(r=(8.53\times10^2)^{1/3}\approx9.5\).
    \item \(m=10^6\), \(a=0.5\): \(r=(8.53\times10^1)^{1/3}\approx4.4\).

\end{itemize}
Thus \(M=10^3\!-\!10^5\,M_\odot\) corresponds to 
\(r\approx9\!-\!44\,R_g\), consistent with inner-disk extents.  
By contrast, \(m\sim10\) yields \(r\approx204\), 
%(i.e.\ \(R\approx204\,R_g\sim6.6\times10^3\) km), 
far outside typical XRB disks \citep{Fragile2007a,Liska2018}, and cannot 
precess rigidly on 44 min timescales.
Therefore, the IMBH regime is the only one that locates a 
rigidly precessing ring at r~$4-40\,R_g$.  However, at the 
small-radius end  (r$\leq5\, R_g$), frame-dragging is so strong 
that the Bardeen-Petterson effect can overcome viscous 
communication, forcing the innermost flow to realign with 
the black hole's equatorial plane and potentially breaking 
the assumption of perfectly rigid precession \citep{Bardeen1975}.

\subsection{Magnetic Field, Power and Emission}

Assume near-Eddington accretion for \(M=10^4\,M_\odot\):
\[ L_{\rm Edd}=1.3\times10^{38}\,\frac{M}{M_\odot}\,\mathrm{erg/s}
=1.3\times10^{42}\,\mathrm{erg/s}, \]
\[ \dot M_{\rm Edd}\approx\frac{L_{\rm Edd}}{0.1\,c^2}
\sim1.4\times10^{21}\,\mathrm{g/s}. \]
At \(r=20\,R_g\), \(R=3.0\times10^{10}\) cm, \(v_{\rm ff}\sim c\), 
the density is
$\rho\sim\frac{\dot M_{\rm Edd}}{4\pi R^2 c}
\approx4.3\times10^{-12}\,\mathrm{g/cm^3}, $
and the ram pressure 
\(P_{\rm ram}\sim\rho\,c^2\approx3.87\times10^9\) dyn/cm\(^2\). 
Equipartition, \(\!B^2/(8\pi)\approx P_{\rm ram}\), gives
\[ B \sim \sqrt{8\pi\,P_{\rm ram}} \sim 3.1\times10^5\,\mathrm{G}, \]
so we adopt \(B\sim10^4\!-\!10^6\) G at \(r\sim10\!-\!40\,R_g\).

%\subsection{Blandford-Znajek Jet Power}

The Blandford-Znajek power is
$P_{\rm BZ}\sim\frac{1}{32}\,a^2\,B^2\,R_g^2\,c$ 
\citep{Blandford1977, Tchekhovskoy2010}.
For \(a=0.5\), \(B=2\times10^5\) G, and \(R_g=1.5\times10^9\) cm,
$ P_{\rm BZ}\approx2.1\times10^{37}\,\mathrm{erg/s}.$
Allowing \(B=10^4\!-\!10^6\) G gives 
\(P_{\rm BZ}\sim10^{35\!-\!39}\) erg/s. 
A radiative efficiency \(\epsilon_j\sim10^{-2}\!-\!10^{-1}\) yields
\[ L_{\rm jet} = \epsilon_j\,P_{\rm BZ} \sim 10^{33\!-\!38}\,\mathrm{erg/s}. \]
In the Appendix ~\ref{appA} we discuss the possibility of 
observing the jet through VLBI.

The combination of a compact emission zone (either at the
jet base or within a narrow magnetospheric gap) and a 
low-density plasma environment minimizes Faraday rotation
and depolarization, allowing intrinsically coherent
radiation to retain a high degree of linear polarization. 
In such regions, the magnetic field is both strong and 
ordered on small scales, naturally favoring maser-like 
or bunching instabilities that can generate bright, 
narrow-band radio bursts.

In addition to the Blandford-Znajek jet, a narrow 
vacuum gap in the black hole magnetosphere could 
also provide a compact, highly ordered emission 
region capable of producing coherent, highly 
polarized bursts. In such a gap the available 
potential drop is
$\Delta_{\rm gap}\approx \frac{\Omega_{\rm BH} B r^2_{\rm g}}
{c}$, where $\Omega_{\rm BH}=a \, c/2r_{\rm g}$ is the angular frequency of the event horizon. The produced electric current
through the Goldreich-Julian characteristic charge density 
$n_{\rm GJ} =\Omega_{\rm BH} \, B/2\pi\, e\, c$, is 
$I_{\rm GJ} \approx n_{\rm GJ} \,e \, c \,  \pi\,  r^2_{\rm g}\sim 
\Omega_{\rm BH}\, B\, r^2_{\rm g}/2c$.
So that the gap power 
$L_{\rm gap}\sim \Delta_{\rm gap} I_{\rm GJ}\sim
10^{37-38} {\rm erg/s}$
for $M\sim 10^4 M_{\odot},$ $B\sim 2\times 10^5$ G, and 
$a=0.5$. Particle-in-cell simulations of black hole gap
discharges show that rapid plasma oscillations can lead to 
charge bunching and pair cascades 
\citep{Levinson2018,Yuan2025}. Thus, whether emerging 
from the jet-launching region or from a magnetospheric 
gap, a compact, ordered field geometry can 
explain the observed coherent, highly polarized radio pulses.

\subsection{Observed Fluxes and Efficiency Constraints}

At an assumed distance \(D = 4.5\)\,kpc (\(\approx1.4\times10^{22}\)cm) 
\citep{Wang2025}, a luminosity \(L_{\rm obs}\) corresponds to flux
$ F_{\rm obs} = \frac{L_{\rm obs}}{4\pi\,D^2}. $
ASKAP observations find peak radio flux densities 
\(S_\nu\sim10\) mJy at \(\nu\sim1.3\) GHz.  
Taking a flat spectrum (\(\alpha\approx0\)), 
the radio luminosity at \(\nu=1.3\) GHz is
$$  L_{\rm radio}\sim4\pi\,D^2\,S_\nu
\approx4\pi\,(1.4\times10^{22})^2\times(10^{-2}\,\mathrm{Jy})
$$ $$
\simeq2.5\times10^{29}\,\mathrm{erg\,s^{-1}\,Hz^{-1}}  $$
or
$ \nu L_\nu\sim3\times10^{38}\,\mathrm{erg/s} $
when integrated over a broad radio band 
(\(\Delta\nu\sim10^9\) Hz).  X-ray follow-ups find
$ F_X\sim10^{-12}\,\mathrm{erg\,cm^{-2}\,s^{-1}}
\quad(0.5\textrm{--}10\,\mathrm{keV}), $
so
\[ L_X\sim4\pi\,D^2\,F_X
\approx4\pi\,(1.4\times10^{22})^2\times10^{-12}
\sim2.5\times10^{34}\,\mathrm{erg/s}. \]
Thus the combined radio + X-ray luminosity during each 
2 min "on"-pulse is \(\sim10^{34\!-\!38}\)\,erg/s.  
Comparing to \(P_{\rm BZ}\sim10^{35\!-\!39}\) erg/s, 
we see that a modest radiative efficiency 
\(\epsilon_j\sim10^{-2}\!-\!10^{-1}\) suffices to 
 power the observed fluxes.
%Concretely, if
% \[ P_{\rm BZ}\approx2\times10^{37}\,\mathrm{erg/s}
% \quad(\text{for }B\approx2\times10^5\mathrm{\,G},\;a=0.5), \]
% then dedicating \(\epsilon_j=10^{-2}\) yields
% \[ L_{\rm jet}\approx2\times10^{35}\,\mathrm{erg/s}, \]
% comfortably above the observed \(2.5\times10^{34}\) 
% erg/s in X-rays alone.  The remaining power can emerge 
% in the radio band.

Demanding
$\epsilon_j \gtrsim \frac{L_{\rm obs}}{P_{\rm BZ}} $
sets a lower limit on \(B\).  If
\(L_X + L_{\rm radio}\sim10^{36}\) erg/s in total, 
and we require \(\epsilon_j\lesssim0.1\), then 
\(P_{\rm BZ}\gtrsim10^{37}\) erg/s, implying 
\(B\gtrsim10^5\) G at \(r\sim20\,R_g\). 
This is fully consistent with our equipartition
estimate \(B\sim10^5\) G.  Conversely, if \(B\) 
were as low as \(10^4\) G, \(P_{\rm BZ}\sim10^{35}\) 
erg/s, and achieving \(L_{\rm obs}\sim10^{36}\) erg/s
would require \(\epsilon_j>1\), which is unphysical.  
Hence we infer
$ B\sim10^5\!-\!10^6\,\mathrm{G} $ as the most plausible range, consistent with 
near-Eddington inflow at \(r\sim10\!-\!30\,R_g\).

Inside the precessing region, the jet-disk system may exhibit 
additional, rapid variability on the local dynamical or 
accretion timescale.  In geometric units, the light-crossing time of \(1\,R_g\) is
$ t_g = \frac{G\,M}{c^3} \simeq 4.9 \times 10^{-6}\,
\mathrm{s}\times\bigl(\tfrac{M}{M_\odot}\bigr).$
For \(M=10^4\,M_\odot\), \(t_g \simeq 4.9\times10^{-2}\) s.  
% A characteristic viscous or accretion timescale at radius \(r\) is
% \[
% t_{\rm acc}(r)\sim \frac{r^2\,t_g}{\alpha_v\,(H/R)^2},
% \]
% where \(\alpha_v\) is the Shakura-Sunyaev viscosity parameter
% and \(H/R\) the disk aspect ratio.  Even adopting a very 
% thick disk (\(H/R\sim0.3\)) and \(\alpha_v\sim0.1\), one finds
% \[
% t_{\rm acc}(20\,R_g)
% \sim \frac{20^2 \times 0.049\,\mathrm{s}}{0.1\times0.3^2}
% \approx \frac{19.6\,\mathrm{s}}{0.009}
% \approx 2.2\times10^3\,\mathrm{s}
% \sim 37\,\mathrm{min}.
% \]
which is very small compared with the 44 min 
precession. 
Also, short-wavelength fluctuations at 
the inner jet footpoint would occur on 
\(\sim10\,t_g\) (\(\sim0.5\!\) s) due to rapid accretion in
tilted disks \citep{Dexter2013} and/or plasmoid 
formation seen in numerical simulations of accreting black holes
\citep{Nathanail2020, Ripperda2020}, most probably
well below any detectable variability.  However, flux eruption events 
which are a prominent feature of strongly magnetized accretion onto a 
black hole, occur every $100-1000$ ${\rm t_g}\approx 5- 50$ s 
\citep{Liska2019a, Ripperda2022, Nathanail2025} and could 
be responsible for the high variability seen in the X-rays.
% Moreover, magnetohydrodynamic 
% "flux eruption events" (e.g.\ \citet{Liska2019a}) can 
% produce quasi-periodic flares every \(\sim1{,}000\,t_g\).  
% For \(M=10^4\,M_\odot\), \(1{,}000\,t_g\approx49\) s-again, 
% far shorter than the 2 min "on" width and completely 
% smeared out by a 2 min integration.  In other words, 
% any variability on \(t\lesssim10^2\) s will be averaged 
% away, leaving only the coherent 44 min precession in the data.

%\quad\Longrightarrow\quad
%m\,r^3 = 1.706\times10^8\,a, 
\section{Discussion}
\label{sec:con}

In this Letter  we invsetigate the possibility that the 
2 min bursts every 44 min from 
ASKAP J1832-0911 can be explained by an intermediate-mass black hole (IMBH) 
and a precessing accretion disk launching a Blandford-Znajek jet, as 
an alternative to magnetar or white-dwarf models.
The Lense-Thirring precession constraint,
$  P_{\rm LT}
= \frac{\pi\,G\,M}{a\,c^3}\,r^3
= 44\ \mathrm{min} $
which for \(a\sim0.3\!-\!0.9\) yields \(r\sim10\!-\!40\,R_g\) 
when \(M\sim10^3\!-\!10^5\,M_\odot\).

IMBH mass-radius solutions of \(M\sim10^3\!-\!10^5\,M_\odot\) 
correspond to \(r\approx9\!-\!44\,R_g\), placing the precessing 
ring well within plausible inner-disk extents.  By contrast, a 
\(\sim10\!-\!30\,M_\odot\) BH would require \(r\sim443\,R_g\), 
which cannot precess rigidly on a 44 min timescale.

% Equipartition arguments at \(r\sim10\!-\!40\,R_g\) imply a magnetic field $ B_{\rm eq}\sim10^5\ \mathrm{G}
% \quad(\text{uncertainty }10^4\!-\!10^6\,\mathrm{G}), $
% assuming near-Eddington accretion for \(M=10^4\,M_\odot\).

% The Blandford-Znajek jet power,
% $P_{\rm BZ}\sim\frac{1}{32}\,a^2\,B^2\,R_g^2\,c
% \sim10^{35\!-\!39}\ \mathrm{erg/s},$
% combined with a modest radiative efficiency 
% \(\epsilon_j\sim10^{-2}\!-\!10^{-1}\), yields
An estimated Blandford-Znajek jet $ L_{\rm jet} \sim10^{33\!-\!38}\ \mathrm{erg/s}, $
easily reproduces the observed \(\sim10^{34\!-\!36}\) 
erg/s in radio + X-rays at \(D\simeq4.5\) kpc.

% Reproducing the observed X-ray flux \(F_X\sim10^{-12}\)\,
% erg cm\(^{-2}\) s\(^{-1}\) (\(L_X\sim2.5\times10^{34}\) erg/s) 
% requires only \(P_{\rm BZ}\sim10^{37}\) erg/s and \(\epsilon_j\sim10^{-2}\), 
% without invoking super-Eddington accretion.

Characteristic inner-disk timescales (\(t_g\sim0.5\) s, MHD 
flux eruptions \(\sim50\) s) can be responsible for the rapid variability in the X-rays in the  2 min integration, and
the 44 min precession remains the only coherent clock.

% Even allowing conical expansion to \(r\sim10^2\,R_g\), the
% jet's angular size \(\lesssim10\,\mu\mathrm{as}\) at 4.5 kpc is 
% below current VLBI resolution (\(\sim200\,\mu\mathrm{as}\)), 
% rendering direct imaging of precession infeasible and leaving 
% this task for future space based VLBI 
% observations.

The main observational tests that uniquely identify a precessing-disk 
IMBH as the engine behind ASKAP J1832-0911, distinguishing it 
from both magnetar and white-dwarf models are the following 
(amore extended discussion on this on Appendix~\ref{appB}. 
All known magnetars spin in the 2-12 s range and exhibit 
large-fraction soft X-ray pulsations, whereas the IMBH jet 
predicts  easily a 2 min / 44 min duty cycle and could potentially 
accommodate some variability due to either accretion or pair cascades.
 Similarly, intermediate-polar white dwarfs display spin/orbital 
 periods of $10^2 - 10^5$ s and characteristic hard bremsstrahlung
 spectra, neither of which are seen in this source. Finally, only 
 an LT-precessing IMBH disk can maintain a stable 2 min / 44 min 
 duty cycle over years without invoking fine-tuned phase locks or state transitions.

\begin{acknowledgements}
The author was supported by the Hellenic Foundation for Research and Innovation (ELIDEK) under Grant No 23698, and by  computational time granted from the National Infrastructures for Research and Technology S.A. (GRNET S.A.) in the National HPC facility - ARIS - under project ID 16033.
\end{acknowledgements}
% WARNING
%-------------------------------------------------------------------
% Please note that we have included the references to the file aa.dem in
% order to compile it, but we ask you to:
%
% - use BibTeX with the regular commands:
%   \bibliographystyle{aa} % style aa.bst
%   \bibliography{Yourfile} % your references Yourfile.bib
%
% - join the .bib files when you upload your source files
%-------------------------------------------------------------------
%----------------------------------------------------------------------
%\section*{Data Availability}
%-----------------------------------------------------------------------
%\smallskip
%
\noindent\textit{\textbf{Data Availability.~}}
The data underlying this article will be shared on reasonable request to the corresponding author.

%\section*{}
\bibliographystyle{aa}
%\begin{thebibliography}
\bibliography{aeireferences, more}

\begin{thebibliography}{49}
\expandafter\ifx\csname natexlab\endcsname\relax\def\natexlab#1{#1}\fi

\bibitem[{{Bardeen} \& {Petterson}(1975)}]{Bardeen1975}
{Bardeen}, J.~M. \& {Petterson}, J.~A. 1975, Astrophys. J., 195, L65

\bibitem[{{Belloni}(2010)}]{Belloni2010}
{Belloni}, T., E., ed. 2010, Lecture Notes in Physics, Berlin Springer Verlag,
  Vol. 794, {The Jet Paradigm}

\bibitem[{Belloni {et~al.}(2000)Belloni, Klein{-}Wolt, Méndez, van~der Klis,
  \& van Paradijs}]{Belloni2000}
Belloni, T.~M., Klein{-}Wolt, M., Méndez, M., van~der Klis, M., \& van
  Paradijs, J. 2000, Astronomy \& Astrophysics, 355, 271

\bibitem[{{Blandford} \& {Znajek}(1977)}]{Blandford1977}
{Blandford}, R.~D. \& {Znajek}, R.~L. 1977, Mon. Not. R. Astron. Soc., 179, 433

\bibitem[{Caproni {et~al.}(2006)Caproni, Livio, Abraham, \&
  Mosquera~Cuesta}]{Caproni2006}
Caproni, A., Livio, M., Abraham, Z., \& Mosquera~Cuesta, H. 2006, Astrophysical
  Journal, 653, 112

\bibitem[{Clarkson {et~al.}(2003)Clarkson, Charles, Coe, Laycock, Tout, \&
  Wilson}]{Clarkson2003}
Clarkson, W.~I., Charles, P.~A., Coe, M.~J., {et~al.} 2003, Monthly Notices of
  the Royal Astronomical Society, 339, 447

\bibitem[{Cooper \& Wadiasingh(2024)}]{Cooper2024}
Cooper, A.~J. \& Wadiasingh, Z. 2024, Monthly Notices of the Royal Astronomical
  Society, 533, 2133

\bibitem[{Corbel {et~al.}(2003)Corbel, Nowak, Fender, Tzioumis, \&
  Markoff}]{Corbel2003}
Corbel, S., Nowak, M.~A., Fender, R.~P., Tzioumis, A.~K., \& Markoff, S. 2003,
  Astronomy \& Astrophysics, 400, 1007

\bibitem[{{Dexter} \& {Fragile}(2013)}]{Dexter2013}
{Dexter}, J. \& {Fragile}, P.~C. 2013, \mnras, 432, 2252

\bibitem[{{Duncan} \& {Thompson}(1992)}]{Duncan1992}
{Duncan}, R.~C. \& {Thompson}, C. 1992, Astrophys. J., 392, L9

\bibitem[{{Enoto} {et~al.}(2017){Enoto}, {Shibata}, {Kitaguchi}, {Suwa},
  {Uchide}, {Nishioka}, {Kisaka}, {Nakano}, {Murakami}, \&
  {Makishima}}]{Enoto2017}
{Enoto}, T., {Shibata}, S., {Kitaguchi}, T., {et~al.} 2017, \apjs, 231, 8

\bibitem[{Farrell {et~al.}(2009)Farrell, Webb, Barret, Godet, \&
  Rodrigues}]{Farrell2009}
Farrell, S.~A., Webb, N.~A., Barret, D., Godet, O., \& Rodrigues, J.~M. 2009,
  Nature, 460, 73

\bibitem[{{Fender} {et~al.}(2004){Fender}, {Belloni}, \& {Gallo}}]{Fender2004}
{Fender}, R.~P., {Belloni}, T.~M., \& {Gallo}, E. 2004, Mon. Not. R. Astron.
  Soc., 355, 1105

\bibitem[{{Fragile} {et~al.}(2007){Fragile}, {Blaes}, {Anninos}, \&
  {Salmonson}}]{Fragile2007a}
{Fragile}, P.~C., {Blaes}, O.~M., {Anninos}, P., \& {Salmonson}, J.~D. 2007,
  \apj, 668, 417

\bibitem[{Gebhardt {et~al.}(2005)Gebhardt, Rich, \& Ho}]{Gebhardt2005}
Gebhardt, K., Rich, R.~M., \& Ho, L.~C. 2005, The Astrophysical Journal, 634,
  1093

\bibitem[{{Harding} \& {Lai}(2006)}]{Harding2006}
{Harding}, A.~K. \& {Lai}, D. 2006, Reports on Progress in Physics, 69, 2631

\bibitem[{Heemskerk \& van Paradijs(1989)}]{Heemskerk1989}
Heemskerk, M.~H.~M. \& van Paradijs, J. 1989, Astronomy \& Astrophysics, 223,
  154

\bibitem[{{Hurley-Walker} {et~al.}(2023){Hurley-Walker}, {Rea}, {McSweeney},
  {Meyers}, {Lenc}, {Heywood}, {Hyman}, {Men}, {Clarke}, {Coti Zelati},
  {Price}, {Horv{\'a}th}, {Galvin}, {Anderson}, {Bahramian}, {Barr}, {Bhat},
  {Caleb}, {Dall'Ora}, {de Martino}, {Giacintucci}, {Morgan}, {Rajwade},
  {Stappers}, \& {Williams}}]{Hurley2023}
{Hurley-Walker}, N., {Rea}, N., {McSweeney}, S.~J., {et~al.} 2023, Nature, 619,
  487

\bibitem[{{Hurley-Walker} {et~al.}(2022){Hurley-Walker}, {Zhang}, {Bahramian},
  {McSweeney}, {O'Doherty}, {Hancock}, {Morgan}, {Anderson}, {Heald}, \&
  {Galvin}}]{Hurley2022}
{Hurley-Walker}, N., {Zhang}, X., {Bahramian}, A., {et~al.} 2022, Nature, 601,
  526

\bibitem[{{Hyman} {et~al.}(2005){Hyman}, {Lazio}, {Kassim}, {Ray}, {Markwardt},
  \& {Yusef-Zadeh}}]{Hyman2005}
{Hyman}, S.~D., {Lazio}, T. J.~W., {Kassim}, N.~E., {et~al.} 2005, Nature, 434,
  50

\bibitem[{Kaspi(2010)}]{Kaspi2010}
Kaspi, V.~M. 2010, Proceedings of the National Academy of Sciences, 107, 7147

\bibitem[{Kaspi \& Beloborodov(2017)}]{Kaspi2017}
Kaspi, V.~M. \& Beloborodov, A.~M. 2017, Annual Review of Astronomy and
  Astrophysics, 55, 261

\bibitem[{Katz(2022)}]{Katz2022}
Katz, J.~I. 2022, Astrophysics and Space Science, 367

\bibitem[{{Levinson} \& {Cerutti}(2018)}]{Levinson2018}
{Levinson}, A. \& {Cerutti}, B. 2018, \aap, 616, A184

\bibitem[{Liska {et~al.}(2018)Liska, Hesp, Tchekhovskoy, Ingram, van~der Klis,
  \& Markoff}]{Liska2018}
Liska, M., Hesp, C., Tchekhovskoy, A., {et~al.} 2018, Mon. Not. R. Astron.
  Soc., 474, L81

\bibitem[{{Liska} {et~al.}(2019){Liska}, {Tchekhovskoy}, {Ingram}, \& {van der
  Klis}}]{Liska2019a}
{Liska}, M., {Tchekhovskoy}, A., {Ingram}, A., \& {van der Klis}, M. 2019,
  \mnras, 487, 550

\bibitem[{Maccarone(2002)}]{Maccarone2002}
Maccarone, T. 2002, Monthly Notices of the Royal Astronomical Society, 336,
  1371

\bibitem[{Makishima {et~al.}(2000)Makishima, Kubota, Mizuno, Ohnishi, Tashiro,
  Aruga, Asai, Dotani, Mitsuda, Ueda, \& Uno}]{Makishima2000}
Makishima, K., Kubota, A., Mizuno, T., {et~al.} 2000, The Astrophysical
  Journal, 535, 632

\bibitem[{Margon(1984)}]{Margon1984}
Margon, B. 1984, Annual Review of Astronomy and Astrophysics, 22, 507

\bibitem[{Merloni {et~al.}(2003)Merloni, Heinz, \& di~Matteo}]{Merloni2003}
Merloni, A., Heinz, S., \& di~Matteo, T. 2003, Monthly Notices of the Royal
  Astronomical Society, 345, 1057

\bibitem[{Mezcua(2017)}]{Mezcua2017}
Mezcua, M. 2017, International Journal of Modern Physics D, 26, 1730021

\bibitem[{{Nathanail} {et~al.}(2020){Nathanail}, {Fromm}, {Porth}, {Olivares},
  {Younsi}, {Mizuno}, \& {Rezzolla}}]{Nathanail2020}
{Nathanail}, A., {Fromm}, C.~M., {Porth}, O., {et~al.} 2020, Mon. Not. R.
  Astron. Soc., 495, 1549

\bibitem[{{Nathanail} {et~al.}(2025){Nathanail}, {Mizuno}, {Contopoulos},
  {Fromm}, {Cruz-Osorio}, {Moriyama}, \& {Rezzolla}}]{Nathanail2025}
{Nathanail}, A., {Mizuno}, Y., {Contopoulos}, I., {et~al.} 2025, \aap, 693, A56

\bibitem[{Patterson(1994)}]{Patterson1994}
Patterson, J. 1994, Publications of the Astronomical Society of the Pacific,
  106, 209

\bibitem[{Petterson(1975)}]{Petterson1975}
Petterson, J.~A. 1975, The Astrophysical Journal, 201, L61

\bibitem[{{Pons} \& {Vigan{\`o}}(2019)}]{Pons2019}
{Pons}, J.~A. \& {Vigan{\`o}}, D. 2019, Living Reviews in Computational
  Astrophysics, 5, 3

\bibitem[{Portegies~Zwart \& McMillan(2002)}]{Portegies_Zwart_2002}
Portegies~Zwart, S.~F. \& McMillan, S. L.~W. 2002, The Astrophysical Journal,
  576, 899–907

\bibitem[{{Qu} \& {Zhang}(2025)}]{Qu2025}
{Qu}, Y. \& {Zhang}, B. 2025, \apj, 981, 34

\bibitem[{Reines \& Volonteri(2015)}]{Reines2015}
Reines, A.~E. \& Volonteri, M. 2015, The Astrophysical Journal, 813, 82

\bibitem[{{Ripperda} {et~al.}(2020){Ripperda}, {Bacchini}, \&
  {Philippov}}]{Ripperda2020}
{Ripperda}, B., {Bacchini}, F., \& {Philippov}, A.~A. 2020, Astrophys. J., 900,
  100

\bibitem[{{Ripperda} {et~al.}(2022){Ripperda}, {Liska}, {Chatterjee}, {Musoke},
  {Philippov}, {Markoff}, {Tchekhovskoy}, \& {Younsi}}]{Ripperda2022}
{Ripperda}, B., {Liska}, M., {Chatterjee}, K., {et~al.} 2022, Astrophys. J.
  Lett., 924, L32

\bibitem[{{Schwope} {et~al.}(2023){Schwope}, {Marsh}, {Standke}, {Pelisoli},
  {Potter}, {Buckley}, {Munday}, \& {Dhillon}}]{Schwope2023}
{Schwope}, A., {Marsh}, T.~R., {Standke}, A., {et~al.} 2023, \aap, 674, L9

\bibitem[{Stella \& Vietri(1998)}]{Stella1998}
Stella, L. \& Vietri, M. 1998, Astrophysical Journal Letters, 492, L59

\bibitem[{{Tchekhovskoy} {et~al.}(2010){Tchekhovskoy}, {Narayan}, \&
  {McKinney}}]{Tchekhovskoy2010}
{Tchekhovskoy}, A., {Narayan}, R., \& {McKinney}, J.~C. 2010, Astrophys. J.,
  711, 50

\bibitem[{{Turolla} {et~al.}(2015){Turolla}, {Zane}, \& {Watts}}]{Turolla2015}
{Turolla}, R., {Zane}, S., \& {Watts}, A.~L. 2015, Reports on Progress in
  Physics, 78, 116901

\bibitem[{Wang {et~al.}(2025)Wang, Rea, Bao, Kaplan, Lenc, Wadiasingh, Hare, \&
  …~Thyagarajan}]{Wang2025}
Wang, Z., Rea, N., Bao, T., {et~al.} 2025, Nature, 619, 487–490

\bibitem[{Warner(1995)}]{Warner1995}
Warner, B. 1995, Cataclysmic Variable Stars (Cambridge University Press)

\bibitem[{{Wilkins}(1972)}]{Wilkins1972}
{Wilkins}, D.~C. 1972, Phys. Rev. D, 5, 814

\bibitem[{{Yuan} {et~al.}(2025){Yuan}, {Chen}, \& {Luepker}}]{Yuan2025}
{Yuan}, Y., {Chen}, A.~Y., \& {Luepker}, M. 2025, \apj, 985, 159

\end{thebibliography}
%\end{thebibliography}
% \newpage
% \newpage

%%%%%%\begin{appendix}
 \appendix
% \section{Table with difference in appearance}
\section{VLBI-Scale Jet Size and Resolvability}

A final test is whether the precessing jet can be spatially
resolved with Very-Long-Baseline Interferometry (VLBI).  
If the jet originates near \(r\approx20\,R_g\), then 
its transverse radius at launch is of order
\[ R_{\rm jet}\sim\alpha\,r\,R_g, \]
where \(\alpha\) is the jet's half-opening angle,
say \(\alpha\simeq10^\circ\approx0.1745\) rad.  Thus
\[ R_{\rm jet}\sim0.1745\times20\times1.5\times10^9\,\mathrm{cm}
\approx5\times10^9\,\mathrm{cm}. \]
Assuming the jet remains collimated on larger scales
but starts with a cylindrical cross-section of radius
\(5\times10^9\) cm, its angular size at \(D=4.5\) kpc is
\[ \theta\sim\frac{5\times10^9}{4.5\times10^{21}}
\approx5.8\times10^{-13}\,\mathrm{rad}
\approx0.12\,\mu\mathrm{as}.  \]
Even if the jet expands conically by a factor of 100
before becoming optically thin 
(i.e.\ radius \(\sim2.6\times10^{11}\) cm), 
\(\theta\sim12\,\mu\mathrm{as}\), still below 
typical VLBI resolution at 1.3 GHz 
(\(\sim200\,\mu\mathrm{as}\)).  Only future 
space-VLBI at higher frequencies could hope to 
resolve any structure.  Therefore, direct imaging 
of the precessing jet is effectively impossible 
with current VLBI; the only observable is 
time-dependent brightness modulation 
("hot spot" lighting up as the beam swings by).
 \label{appA}

\section{Discriminating IMBH, Magnetar and White-Dwarf Scenarios}
\label{appB}

In this section, we summarize the "smoking-gun" 
tests that distinguish a precessing-disk IMBH from 
alternative magnetar or white-dwarf models, and 
explain why a stellar-mass (\(\sim10\!-\!30\,M_\odot\))
black-hole X-ray binary (XRB) is effectively ruled 
out-short of observing a state transition lasting 
months-years that temporarily erases and then restores 
the 2 min / 44 min pattern. 
 % In the Appendix~\ref{appB} we discuss why a  Low-Mass 
 % XRB is mostly not favored.

\subsection{Magnetar Model Versus IMBH Jet}

All known magnetars spin with 
\(P_{\rm spin}\sim2\!-\!12\) s and exhibit coherent 
pulsations in soft X-rays (0.5-10 keV) with large 
pulsed fractions (\(\sim10\!-\!50\%\)) \citep[e.g.,][]{Pons2019,Kaspi2017}.  
A slowly rotating (\(\sim44\) min) magnetar would spin
down catastrophically fast
(\(\dot P\sim10^{-9}\!-\!10^{-8}\,\mathrm{s/s}\) for 
\(B\sim10^{14}\!-\!10^{15}\) G), making long-term stability 
impossible \citep{Duncan1992,Harding2006}.  By contrast, an
IMBH jet model predicts  \(\sim\)s-scale variability; 
in the 2 min precession on window and, at 
most, high-frequency flickering on \(\lesssim10^2\) s timescales
\citep{Liska2018}.  Deep, high-time-resolution X-ray timing 
can rule out any \(P_{\rm spin}<12\) s.
% Non-detection of coherent \(\sim\)s-pulses during the 2 min "on" window 
% would falsify a magnetar interpretation.

Persistent magnetar spectra are well modeled by one or two 
blackbodies (\(kT\sim0.3\!-\!0.6\) keV) plus a hard power-law
tail (\(\Gamma\sim2\!-\!4\)) \citep{Enoto2017,Turolla2015},
with no cool (\(<0.1\) keV) disk component.  An IMBH, however,
should show a multicolor disk blackbody peaking at 
\(kT_{\rm in}\sim0.05\!-\!0.1\) keV (for
\(M\sim10^4\,M_\odot\)) plus a nonthermal tail from 
jet-disk coupling \citep{Makishima2000,Farrell2009}. 
% Fitting the 0.3-10 keV spectrum to detect a cool disk 
% component (\(<0.1\) keV) with \(L_{\rm disk}\propto T_{\rm in}^4\) 
% scaling would confirm an accretion disk around a \(\gtrsim10^3\,M_\odot\)
% BH; a pure magnetar-like blackbody + steep PL would argue against the IMBH model. % \citep{Pons2019}.

% When radio-active, magnetars emit short (\(<1\) s) pulses 
% with steep spectra (\(\alpha\sim-1.5\)) and nearly 100\% polarization 
% \citep{Camilo2006,Serylak2009}.  They do not produce 
% \(\sim2\) min continuous flares on a 44 min cycle. 

% A compact
% IMBH jet instead exhibits a flat or mildly inverted spectrum 
% (\(\alpha\sim0\!-\!0.2\)) and modest polarization (\(\lesssim10-20\%\)),
% with a smooth swing in polarization angle as the jet precesses 
% \citep{Broderick2009,Nemmen2007}.  Multi-frequency radio 
% observations during the 2 min "on" intervals can measure
% \(\alpha_\nu\) and polarization fraction: a flat spectrum 
% and \(\le10\%\) polarization supports a synchrotron jet, while a
% steep, highly polarized burst would be magnetar-like.

Magnetars are typically isolated or in young supernova 
remnants/OB associations \citep{Kaspi2010}.  A bright optical/IR
counterpart showing magnetospheric pulsations (\(\sim\)s)  would 
favor a magnetar.  In contrast, an IMBH in a dwarf galaxy or 
globular cluster may be optically faint (\(M_V>-2\)) or 
undetected at 4.5 kpc \citep{Mezcua2017}.  Deep HST 
or 8 m-class imaging: detection of an OB companion or SNR shell 
implies a magnetar; absence of a stellar counterpart, possibly a 
faint extended cluster, favors the IMBH.

\subsection{White-Dwarf Model Versus IMBH Jet}

Intermediate polar white dwarfs have \(P_{\rm spin}\sim10^2\!-\!10^3\) s
and orbital periods \(P_{\rm orb}\sim10^4\!-\!10^5\) s.  A beat between 
spin and orbital cannot naturally produce a strict 44 min 
(\(\sim2640\) s) period without fine tuning 
\citep{Patterson1994, Qu2025}. 
Such systems would also exhibit coherent pulsations at the white-dwarf 
spin (\(\sim500\) s) or orbital (\(\sim10^4\) s) periods in X-ray/optical 
light curves \citep{Warner1995}.  An IMBH jet model predicts
none of these intermediate periods-only the 44 min precession and 
high-frequency flicker (\(<10^2\) s).  Searching for \(10^2\!-\!10^4\) s
pulsations can thus discriminate a white-dwarf origin.

% Accreting magnetic white dwarfs produce hard bremsstrahlung
% (\(kT\sim10\!-\!20\) keV) with Fe K\(\alpha\) lines and sometimes 
% optical/IR cyclotron harmonics \citep{Ezuka1999,Ramsay2004}.  An 
% IMBH disk shows a cool (\(<0.1\) keV) blackbody  + PL tail; detection 
% of Fe line complexes or high-temperature bremsstrahlung strongly 
% points to a white dwarf, while a cool disk plus PL tail rules it out.

% Some polars and IPs show sporadic cm-wave flares via cyclotron 
% maser or gyrosynchrotron, with steep spectra (\(\alpha<-1\)) 
% and high circular polarization (\(>50\%\)) \citep{Coe2004,Coppejans2015}. 
% These last minutes-hours but are not locked to 44 min.  
% An IMBH jet produces continuous synchrotron flares timed by 
% precession, with moderate linear polarization and a flat spectrum.
% Measuring circular vs.\ linear polarization and spectral slope can 
% thus distinguish cyclotron-maser white-dwarf flares from an IMBH jet.

Achieving a stable \(\Delta t/P=2/44\) min duty cycle over
months would require an implausible long-term phase lock 
between spin and orbit in a white-dwarf system \citep{Warner1995},
as tiny changes in \(\dot P\) or mass-transfer rate would break 
the 44  min clock.  An LT-precessing IMBH disk is expected to 
maintain \(P_{\rm LT}\) stably for \(10^4\!-\!10^6\,t_g\) 
\citep{Fragile2007a,Liska2018}.  Monitoring the period over
\(\gtrsim6\) months: any \(\Delta P/P\gtrsim1\%\) drift is 
inconsistent with LT precession but expected for white-dwarf
spin evolution (\(\dot P\sim10^{-11}\) s/s).
 \subsection{Low-Mass XRB Versus IMBH}

For a stellar-mass BH (\(m\sim10\)), Eq.~\eqref{eq:mass_radius}
demands \(r\sim443\,R_g\) (\(R\sim6.6\times10^3\) km) to satisfy
\(P_{\rm LT}=44\) min.  At such radii, the viscous timescale
\(t_{\rm acc}\) is \(\sim11\) days \citep{Fragile2007a,Liska2019a},
preventing rigid precession.  Known XRBs precess on much longer 
superorbital timescales (SS 433: 162 d; Her X-1: 35 d; LMC X-4: 30 d; SMC X-1: 60 d) \citep{Margon1984,Petterson1975,Heemskerk1989,Clarkson2003}.  
Thus \(10\!-\!30\,M_\odot\) XRBs cannot explain a 44 min precession.

Moreover, a ten-solar-mass XRB jet at \(r\sim443\,R_g\) would
have too low \(P_{\rm BZ}\) (\(\sim10^{31\!-\!33}\) erg/s) to 
power the observed \(L_X\sim10^{34\!-\!36}\) erg/s \citep{Merloni2003}. 
In canonical XRB behavior, a transition from the hard state 
(jet "on") to soft state (jet quenched) occurs over hours-days 
\citep{Fender2004,Corbel2003}, during which the radio flares vanish 
and the X-ray spectrum softens.  Only if ASKAP J1832-0911 were to 
undergo such a state transition-suppressing the 2/44 pattern for
days-weeks and later restoring it-could a low-mass XRB remain 
viable, a behavior not yet observed in this source.

In canonical XRB behavior, a transition from the low/hard state 
(jet active, hard X-ray power law) to the high/soft state (jet 
quenched, disk-dominated spectrum) occurs over hours-days
\citep{Fender2004,Corbel2003}.  During this transition, the 
flat-spectrum radio core disappears and the nonthermal X-ray 
tail softens \citep{Belloni2010}.  If ASKAP J1832-0911 were a 
ten-solar-mass XRB, the 2 min/44 min flares would persist until
such a state transition, vanish for days-weeks, and only 
reappear after returning to the hard state over months-years
(as seen in GRS 1915+105; \citet{Belloni2000}).  No prolonged
"off-period" has been observed to date.  In the absence of any
detected state transition, a \(10\!-\!30\,M_\odot\) XRB
interpretation is strongly disfavored.  Only a future 
detection of jet quenching and spectral softening lasting days, 
followed by a re-establishment of the 2/44 pattern after years, 
could salvage the low-mass XRB scenario.

\end{document}